\documentclass[12pt]{iopart}

\usepackage{lscape}
\newcommand{\unit}[1]{~\mathrm{#1}}

\newcommand{\zu}[1]{Fig.~{\ref{#1}}}
\newcommand{\hyou}[1]{Table~{\ref{#1}}}
\newcommand{\Ref}[1]{Ref.~{\cite{#1}}}
\newcommand{\Sec}[1]{Sec.~{\ref{#1}}}
\def\rtHz{\sqrt{Hz}}

\usepackage{graphicx} 
\usepackage{hyperref}

\begin{document}
\title[Mirror actuation design for interferometer control of KAGRA]{Mirror actuation design for the interferometer control of the KAGRA gravitational wave telescope}

\author{Yuta Michimura$^1$, Tomofumi Shimoda$^1$, Takahiro Miyamoto$^2$, Ayaka Shoda$^3$, Koki Okutomi$^{3,4}$, Yoshinori Fujii$^3$, Hiroki Tanaka$^2$, Mark A. Barton$^3$, Ryutaro Takahashi$^3$, Yoichi Aso$^{3,4}$, %
Tomotada Akutsu$^3$, Masaki Ando$^{1,3,5}$, Yutaro Enomoto$^1$, Raffaele Flaminio$^3$, Kazuhiro Hayama$^6$, Eiichi Hirose$^{2,6}$, Yuki Inoue$^{7,8}$, Takaaki Kajita$^2$, Masahiro Kamiizumi$^6$, Seiji Kawamura$^6$, Keiko Kokeyama$^6$, Kentaro Komori$^1$, Rahul Kumar$^8$, Osamu Miyakawa$^6$, Koji Nagano$^2$, Masayuki Nakano$^2$, Naoko Ohishi$^3$, Ching Pin Ooi$^1$, Fabi{\'a}n Erasmo Pe{\~n}a Arellano$^3$, Yoshio Saito$^6$, Katsuhiko Shimode$^6$, Kentaro Somiya$^9$, Hiroki Takeda$^1$, Takayuki Tomaru$^8$, Takashi Uchiyama$^6$, Takafumi Ushiba$^2$, Kazuhiro Yamamoto$^{10}$, Takaaki Yokozawa$^{11}$, Hirotaka Yuzurihara$^{11}$}
\address{$^1$ Department of Physics, University of Tokyo, Bunkyo, Tokyo 113-0033, Japan}
\address{$^2$ Institute for Cosmic Ray Research, University of Tokyo, Kashiwa, Chiba, 277-8582, Japan}
\address{$^3$ National Astronomical Observatory of Japan, Mitaka, Tokyo, 181-8588, Japan}
\address{$^4$ SOKENDAI, The Graduate University for Advanced Studies, Hayama, Kanagawa 240-0193, Japan}
\address{$^5$ Research Center for the Early Universe, University of Tokyo, Bunkyo, Tokyo 113-0033, Japan}
\address{$^6$ KAGRA Observatory, Institute for Cosmic Ray Research, University of Tokyo, Hida, Gifu 506-1205, Japan}
\address{$^7$ Institute of Physics, Academia Sinica, Nankang, Taipei 11529, Taiwan}
\address{$^8$ High Energy Accelerator Research Organization, Tsukuba, Ibaraki, 305-0801, Japan}
\address{$^9$ Department of Physics, Tokyo Institute of Technology, Ookayama, Meguro, Tokyo 152-8550, Japan}
\address{$^{10}$ Department of Physics, University of Toyama, Toyama, Toyama, 930-8555, Japan}
\address{$^{11}$ Department of Physics, Osaka City University, Sumiyoshi, Osaka 558-8585, Japan}

\ead{michimura@granite.phys.s.u-tokyo.ac.jp}
\vspace{10pt}
\begin{indented}
\item[]\today
\end{indented}

\begin{abstract}
KAGRA is a 3-km cryogenic interferometric gravitational wave telescope located at an underground site in Japan. In order to achieve its target sensitivity, the relative positions of the mirrors of the interferometer must be finely adjusted with attached actuators. We have developed a model to simulate the length control loops of the KAGRA interferometer with realistic suspension responses and various noises for mirror actuation. Using our model, we have designed the actuation parameters to have sufficient force range to acquire lock as well as to control all the length degrees of freedom without introducing excess noise.
\end{abstract}


\vspace{2pc}

\section{Introduction}
The discovery of gravitational waves by Advanced LIGO has opened a brand new window to our Universe~\cite{GW150914}. To further enhance gravitational wave astronomy with better sky localization, better sky coverage, and more precise parameter estimation~\cite{ObservationScenario,SchutzNetwork}, it is essential to extend the global network of advanced gravitational wave telescopes, with detectors such as Advanced Virgo~\cite{AdVirgo}, KAGRA~\cite{SomiyaKAGRA,AsoKAGRA,AkutsuKAGRA}, and the third LIGO detector in India~\cite{LIGOIndia}.

KAGRA is a 3-km interferometric gravitational wave telescope located in the Kamioka mine in Gifu Prefecture, Japan. Two unique features of KAGRA among advanced interferometers are that it is constructed at a seismically quiet underground site~\cite{UchiyamaTunnel}, and that it uses sapphire mirrors at cryogenic temperature to reduce thermal noise~\cite{HiroseSapphire,HiroseCoating}. KAGRA was funded in 2010 and the tunnel excavation was started in May 2012, which continued until the end of March 2014. In March 2016 the initial phase of KAGRA, with a simple 3-km Michelson interferometer configuration, was operated to test the basic performance at room temperature~\cite{iKAGRA}.

To operate an interferometric gravitational wave telescope with high sensitivity, the relative positions of the mirrors must be finely tuned to maintain the resonant conditions of the cavities. For this purpose, actuators are attached to the mirrors and various locations in their suspension systems. Those actuators should have a wide enough force range to keep the relative motion of the mirrors small enough, without introducing excess noise into the gravitational wave signal. Minimizing the noise coming from the actuators is essential for achieving the designed sensitivity, especially in the lower end of the detector bandwidth, at a few tens of Hertz.

In this paper, we show noise and range calculations for designing the mirror actuation used for the length control of the KAGRA interferometer. We start by describing the overall interferometer configuration and the mirror suspension configuration of KAGRA. We then describe our simulation model used to calculate the noise and the range of the actuators. Finally, we present the results of our simulation to show that our actuator design meets the displacement noise requirements for each mirror, and has a wide enough force range for the length control of the KAGRA interferometer. We also discuss the force range required for the lock acquisition of the interferometer.

\section{Interferometer configuration} \label{Sec:IFO}

\begin{figure}
	\begin{center}
		\includegraphics[width=12cm]{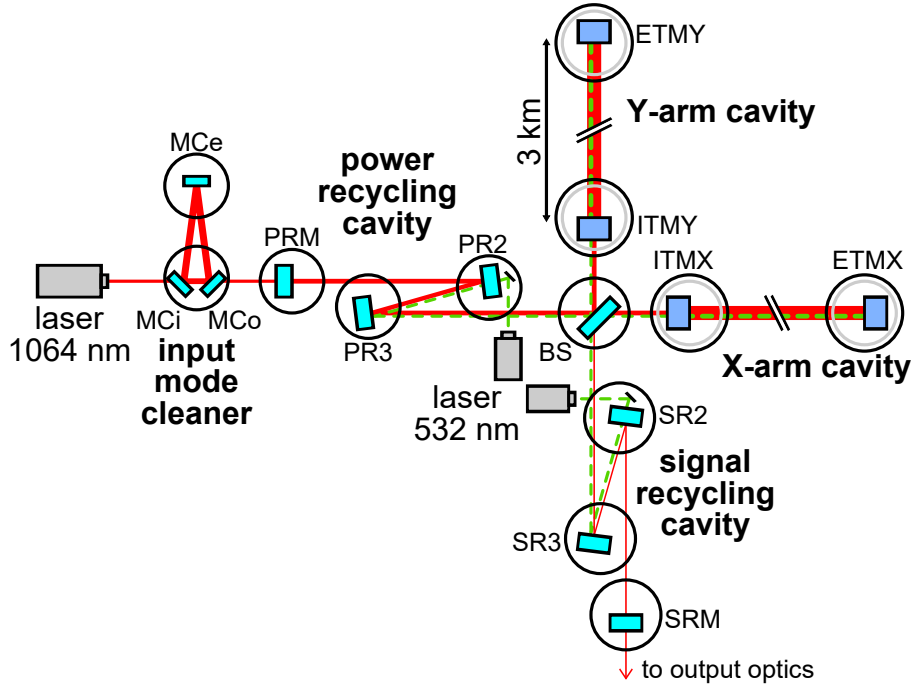}
	\caption{Schematic of the KAGRA interferometer. All the mirrors and the beam paths of the input mode cleaner and the main interferometer are enclosed in the vacuum system. ITMs and ETMs are cryogenic sapphire mirrors, and all the other mirrors are fused silica mirrors at room temperature. Frequency-doubled auxiliary laser beams are injected from PR2 and SR2 for arm-length stabilization.}
	\label{IFOConfig}
	\end{center}
\end{figure}

The main part of the KAGRA interferometer is a resonant sideband extraction interferometer~\cite{MizunoRSE} formed by suspended mirrors. The schematic of the KAGRA interferometer is shown in \zu{IFOConfig}. We use a single-frequency continuous-wave laser source with a wavelength of 1064~nm. The laser beam first passes through a triangular ring cavity, called the input mode cleaner (IMC), to clean the spatial modes of the laser before reaching the main interferometer. The IMC consists of three mirrors (MCi, MCo, and MCe), and its round-trip cavity length and finesse are 53.3~m and 540, respectively. The IMC length (IMCL) control is done by adjusting the position of MCe and the frequency of the laser. Since the length of the IMC, formed by suspended mirrors, is stable above the pendulum resonant frequency, the laser frequency follows the IMC resonant frequency above $\sim 1\unit{Hz}$ for pre-stabilization~\cite{aLIGOPSL}.

The main interferometer has two 3-km long arm cavities formed by cryogenic sapphire input test masses (ITMs) and end test masses (ETMs). The reflected beams from the arm cavities are combined at a beam splitter (BS) to form a Michelson interferometer. A power-recycling mirror (PRM) and the two ITMs together with PR2 and PR3 mirrors form a folded power-recycling cavity (PRC) to effectively increase the input power. Similarly, a signal-recyling mirror (SRM) and the two ITMs together with SR2 and SR3 mirrors form a folded signal-recycling cavity (SRC). The SRC length is tuned so that the storage time of the gravitational wave signal in the arm cavities is reduced to broaden the detector bandwidth. This scheme is called resonant sideband extraction.

The input laser power to the main interferometer at the back of PRM is 78~W, and the power recycling gain is 10. The arm cavity finesse is 1530 and the SRM transmission is 15\%. These parameters are related to the spectral shape of the quantum noise, and are chosen to maximize the distance at which we can detect gravitational wave signals from neutron star binaries~\cite{SomiyaKAGRA}.
There is an option to detune the SRC for further optimization. In this paper, we only show the result of the calculation for the non-detuned (broadband) case, but a very similar result is obtained in the detuned case. A more detailed description of the main interferometer and its parameters is given in \Ref{AsoKAGRA}.

We have five degrees of freedom for the main interferometer length control. The differential length change of the arm cavities, which is called DARM (see \hyou{LengthUGF}), is the most important because signals from gravitational waves appear in DARM. DARM is controlled by actuating ETMX and ETMY differentially. The common length change of the arm cavities, CARM, is used as an ultimate reference for the laser frequency stabilization, and is controlled by a combination of laser frequency actuators. Details of the frequency stabilization system, including the CARM loop, are discussed in~\Ref{KAGRAFSS}. The differential length change of the Michelson interferometer, MICH, is controlled via the BS. The PRC and SRC lengths, PRCL and SRCL, are controlled via PRM and SRM, respectively. See \hyou{LengthUGF} in a later section for a summary of the properties of the length control loops considered in this paper.

The lock acquisition of all the five length degrees of freedom simultaneously is difficult because the length error signals are highly nonlinear and highly coupled to each other. Therefore, to acquire the lock of the main interferometer robustly, we use the arm-length stabilization system~\cite{StaleyALS}. 532~nm beams from frequency-doubled auxiliary laser sources are injected into both arms from the back of PR2 and SR2. The arm cavity finesse for 532~nm is designed to be 50 for easier lock acquisition than the main 1064~nm beam.

First, the arm cavities are locked to these auxiliary lasers, which are phase-locked to the main laser with some offset. In this way the arm cavity lengths are stabilized off resonance with respect to the main laser, and the central dual-recycled Michelson part of the main interferometer can be locked without disturbance from the length change of the arm cavities. After locking the central part, we then bring the arm cavities to resonance with the main laser by changing the offset frequency of the auxiliary lasers to the main laser. Finally, we switch the error signals for the arm cavities from the ones from the auxiliary lasers to the ones from the main laser.

After bringing all the degrees of freedom to the operation points with the main laser, the arm-length stabilization system with auxiliary lasers is turned off. Although this system is only used for the lock acquisition, it is important to take it into account in the actuator range calculation because length sensing noise with auxiliary laser is worse than that with the main laser. The auxiliary laser has lower power and thus higher shot noise, as discussed in \Sec{Sec:Quantum}.

\section{Suspension configurations}

\begin{figure}
	\begin{center}
		\includegraphics[width=15cm]{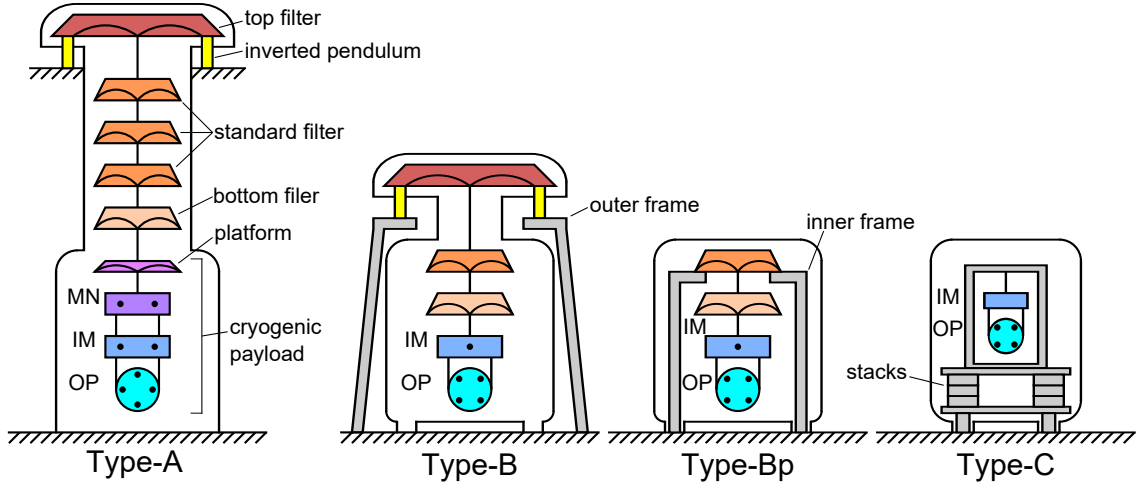}
	\caption{Schematic of the vibration isolation systems. MN: marionette, IM: intermediate mass, OP: optic. Black dots represent the location of coil-magnet actuators. Type-A systems are located inside a vacuum tower and the legs of the inverted pendulum table is fixed onto the ground of the second floor. The legs of the inverted pendulum table of the Type-B system are fixed onto the outer frame of the vacuum chamber. The standard filter of the Type-Bp system are fixed onto the inner frame of the vacuum chamber. Type-C systems are fixed on the vibration-isolated table with 3-stage stacks.}
	\label{SuspensionSystems}
	\end{center}
\end{figure}

The mirrors for the KAGRA interferometer are suspended from different types of vibration isolation systems depending on their displacement noise requirements. Mirrors with stringent displacement noise requirements are suspended by vibration isolation systems with multiple stages to attenuate the noise from ground motion. The schematic of the vibration isolation systems used for KAGRA is shown in \zu{SuspensionSystems}. ITMs and ETMs are suspended by a 14-m long eight-stage pendulum called Type-A~\cite{TypeA}. The pendulum is hung from a pre-isolator which consists of a vertical geometric anti-spring (GAS) top filter~\cite{GASFilter} supported by an inverted pendulum table~\cite{TakamoriIP}. The Type-A system extends over two stories and the legs of the inverted pendulum table are fixed onto the second floor. The inverted pendulum table is also used for adjusting the position and the alignment of the whole suspension chain. From the top GAS filter, three standard GAS filters, a bottom GAS filter, platform, marionette (MN), intermediate mass (IM), and the optic (OP) are suspended in the order as mentioned.

\begin{landscape}
\begin{table}
  \begin{center}
    \caption{KAGRA suspension and actuator parameters. The coil-magnet actuation efficiencies for Type-A and Type-C are measured values. All the other parameters are designed values.} \label{ActuatorParameters}
    \begin{tabular}{lccccc}
      \br
Type  &   Type-A  &   Type-B (BS)  &   Type-B (SR)  &   Type-Bp  &   Type-C   \\
\hline
Mirrors   &   ITM/ETM   &   BS   &   SRM/SR2/SR3   &   PRM/PR2/PR3   &   MCi/MCo/MCe   \\
Mirror diameter [mm]   &   220   &   370   &   250   &   250   &   96   \\
Mirror thickness [mm]   &   150   &   80   &   100   &   100   &   30   \\
Mirror mass [kg]   &   22.8   &   18.9   &   10.8   &   10.8   &   0.47   \\
IM (MN) mass [kg]   &   20.5 (22.5)   &   36.5   &   15.6   &   15.6   &   0.71   \\
Wire length for OP [m]  &   0.35   &   0.5   &   0.5   &   0.5   &   0.25   \\
Wire length for IM (MN) [m]  &   0.26 (0.35)   &   0.5   &   0.5   &   0.5   &   0.25   \\
\hline
OP coil turns  & 100 & 600 & 600 &  600  &  41  \\
OP coil resistance [$\Omega$]  & 0.6 & 12 & 12 &  12  &  2.5  \\
OP magnet size [mm]   &   $\phi 2 \times 2 \rm{t}$   &   $\phi 2 \times 3 \rm{t}$   &   $\phi 2 \times 5 \rm{t}$   &   $\phi 6 \times 3 \rm{t}$   &   $\phi 1 \times 5 \rm{t}$   \\
OP actuation per coil [N/A]   &  0.0015   & 0.014 & 0.023 &  0.13  &   0.0014   \\
\# of OP longitudinal coils   & 4 & 4 & 4 &  4  &  4  \\
OP coil driver type   &  Low  &  Low  &  Low  &  High  &  High  \\
\hline
IM coil turns  & 600 & 600 & 600 &  600  &    \\
IM coil resistance [$\Omega$]  & 2 & 12 & 12 &  12  &   \\
IM magnet size [mm]   &   $\phi 2 \times 2 \rm{t}$   &   $\phi 10 \times 10 \rm{t}$   &   $\phi 10 \times 10 \rm{t}$   &   $\phi 10 \times 10 \rm{t}$   &      \\
IM actuation per coil [N/A]   & 0.016 & 1.3 & 1.3 &  1.3  &      \\
\# of IM longitudinal coils   & 2 & 1 & 1 &  1  &    \\
IM coil driver type   &  Modified low  &  Low  &  Low  &  High  &    \\
\hline
MN coil turns  & 600 &  &  &  &    \\
MN coil resistance [$\Omega$]  & 2 &  &  &    &   \\
MN magnet size [mm]   &   $\phi 5 \times 13 \rm{t}$   &  &  &  &      \\
MN actuation per coil [N/A]   & 0.43 &  &  &  &      \\
\# of MN longitudinal coils   & 2 &  &  &  &    \\
MN coil driver type   &  Modified low  &  &  &  &    \\
      \br
    \end{tabular}
  \end{center}
\end{table}
\end{landscape}

The last four stages of the Type-A system are cooled down to cryogenic temperatures and are called the cryogenic payload~\cite{Cryopayload}. The sapphire mirror is cooled down to 20 K and the other parts of the crygenic payload are cooled down to 16 K. The mirror has a higher temperature than the rest because of the heat absorption from the laser beam. The heat absorbed in the mirror is extracted via 4 sapphire fibers which hang the mirror from the IM~\cite{SapphireFiber}. From the IM, the heat is transferred through pure aluminum flexible wires connected to upper stages and then to cryocoolers.

For the room-temperature fused-silica mirrors, simpler systems are used. BS, SRM, SR2 and SR3 are each suspended by a four-stage pendulum called Type-B~\cite{FabianRSI,SekiguchiPhD}. Type-B systems consist of an inverted pendulum table, a top GAS filter, a standard GAS filter, a bottom GAS filter, an IM, and an OP. PRM, PR2 and PR3 are each suspended from a triple pendulum called Type-Bp, which is a simplified version of Type-B. Instead of the inverted pendulum table, the Type-Bp system is supported by a set of motorized linear stages, called a traverser, for adjusting the position and the alignment of the chain in the horizontal plane. IMC mirrors are each suspended from a double pendulum fixed on a three-stage vibration isolation stack~\cite{TakahashiStack}. This system is called a Type-C system and is a modified version of the suspension used for the TAMA300 gravitational wave detector~\cite{TypeC}.

Various kinds of actuators are integrated in the suspension systems for position and alignment controls of the pendulum. The position and alignment controls include resonant mode damping servos using local displacement sensors integrated in the suspension systems~\cite{FabianRSI}, and global controls using the interferometer error signals. Here, we focus on the global length control of the interferometer using longitudinal actuators. There are also vertical and translational actuators, but their effect on the length control is negligible.

\begin{figure}
	\begin{center}
		\includegraphics[width=8cm]{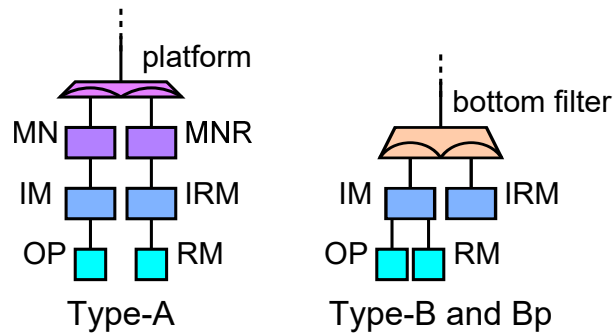}
	\caption{Diagram of the recoil mass chain for the Type-A, Type-B and Bp systems. MNR: marionette recoil mass, IRM: intermediate recoil mass, RM: recoil mass.}
	\label{RecoilMass}
	\end{center}
\end{figure}

The longitudinal actuators used for the length control consist of coils and magnets, and actuation is done by controlling the current applied to the coils. The magnets are glued onto the MN, IM and OP, and coils are fixed on their respective recoil masses or suspension frames (see \zu{RecoilMass}). For Type-A suspension, the recoil mass chain is suspended from the platform, independently from the main optic chain. For Type-B and Type-Bp suspensions, the recoil mass for the IM is suspended from the bottom GAS filter, with the recoil mass for the OP suspended from the IM. Type-C suspensions do not have actuators on the IM stage, and OP coils are fixed on the suspension frame.

\begin{table}
  \begin{center}
    \caption{Operational amplifiers used for the current source, output resistances at DC of the coil drivers. The maximum output current applied to a coil from each coil driver with DAC output of 10~V is also shown. The resistance of the coils is not included in the output resistance values. The maximum output current of the low power coil drivers for Type-A OP is 1.3~mA, and that for others are 1.7~mA.} \label{CoilDrivers}
    \begin{tabular}{lccccc}
      \br
     & Current source & Output resistance & Max current to coil \\
\hline
High power        & OPA548 & $80\unit{\Omega}$ & 0.12~A \\
Low power         & AD8671 & $7.8\unit{k \Omega}$ & 1.7~mA or 1.3~mA \\
Modified low power& AD8671 & $1.4\unit{k \Omega}$ & 9.5~mA \\
      \br
    \end{tabular}
  \end{center}
\end{table}

\hyou{ActuatorParameters} summarizes the actuator parameters for each suspension. Mirror sizes and suspension wire lengths are determined from vibration isolation requirements and other geometric reasons. Numbers of turns for coils, magnet sizes and coil driver parameters are determined based on noise and range calculations described in the following sections. The current applied to the coils is controlled from a digital system. The KAGRA digital system~\cite{DGS} has 16 bit digital-to-analog converters (DACs) with a range of $\pm 10\unit{V}$. We use three types of coil drivers; a high power coil driver, low power coil driver, and modified version of the low power coil driver. The low power coil driver has a pole at 50~Hz (160~Hz for modified version), and a zero at 310~Hz. It also has a voltage gain of 1.33, except for the one used for the OP stage of the Type-A suspension. Operational amplifiers used for the current source are different for each driver so that the maximum output current is defined by the DAC limit, and not the coil driver. The design of each driver is summarized in \hyou{CoilDrivers}.

The maximum current we can apply to the coil considering the damage to the coil is more than 500~mA for Type-C, and 100~mA for other suspensions, which is above the DAC limit. The heat generated by the coils for the Type-A suspension is calculated to be 2.2~mW at maximum, which is sufficiently small compared with the heat extraction capability of the cryogenic system~\cite{SakakibaraCryo}.

The coil-magnet actuation efficiencies in units of N/A in \hyou{ActuatorParameters} for Type-A and Type-C are measured values, and those for Type-B and Type-Bp the calculated values~\cite{MarkOSEM}. The actuation efficiencies depend on the position of the magnet relative to the coil, and the relative position is set to maximize the efficiency within 1~mm. The positional deviation by 1~mm from the optimal point decreases the efficiency by approximately 5\%.
The efficiency could also vary depending on the variation of the magnetic moment of each magnet, which is measured to be smaller than 20\%.

Except for the Type-C suspension, we use samarium-cobalt (SmCo) magnets which are less affected by Barkhausen noise, instead of neodymium (NdFeB) magnets which were used in Initial LIGO~\cite{LIGOBarkhausen}. For Type-C suspensions, we use NdFeB magnets, since their displacement noise requirement is less stringent than the other suspensions.

\section{Length control loop model}

\begin{figure}
	\begin{center}
		\includegraphics[width=12cm]{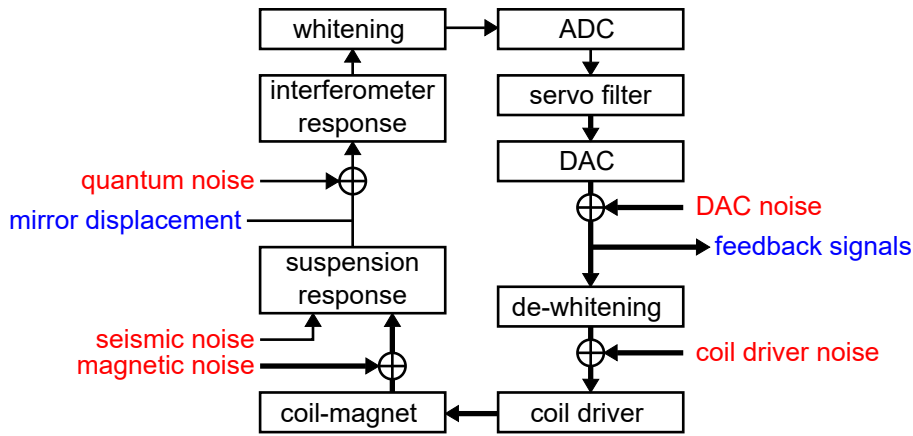}
	\caption{Architecture of the length control loop model to calculate the noise and range of the mirror actuators. Anti-aliasing and anti-imaging filters are not shown for simplicity. The servo filter generates the feedback signals for each suspension stage using the mirror displacement signal.}
	\label{LoopModel}
	\end{center}
\end{figure}

To calculate the noise from the actuators and to calculate the force range needed for the length control, we have developed a length control loop model. We used a MATLAB Simulink-based frequency-domain noise calculation tool, called SimulinkNB~\cite{SimulinkNB} for this modeling. SimulinkNB parses a block diagram made with MATLAB Simulink, and calculates the spectrum of a signal at a specified point of the diagram using transfer functions and noises in the diagram. 

The diagram of our model is shown in \zu{LoopModel}. The interferometer response from the mirror displacements to the length error signals extracted from photo-diodes at interferometer detection ports are calculated using a frequency-domain interferometer simulation tool, called Optickle~\cite{Optickle}. The DARM, MICH, and IMCL error signals are extracted from the transmitted beam from SRM, the reflected beam from PRM, and the reflected beam from MCi, respectively. The PRCL and SRCL error signals are extracted from the transmitted beam of PR2, which is a pick-off beam from the power reycling cavity. Details of the simulation results of the KAGRA interferometer length signal extraction are addressed in \Ref{AsoLSC}.

The length error signals are sent into the KAGRA digital system via 16-bit analog-to-digital converters (ADCs). ADCs and DACs operate at a sampling frequency of 65536~Hz, and the signals are down-converted to 16384~Hz by the digital system. We use 3rd-order Butterworth low-pass filters with a cut-off frequency at 10~kHz for anti-aliasing and anti-imaging.

The digital system generates the feedback signal for each stage of the suspension with infinite impulse response servo filters. The feedback signals are then converted into analog signals by DACs, and sent to coil drivers for each suspension stage. Whitening filters and de-whitening filters are inserted before ADCs and after DACs respectively to effectively reduce ADC and DAC noise. Transfer functions from the actuator force applied to each suspension stage to the displacement of the mirror are calculated using the suspension rigid-body simulation tool, called SUMCON~\cite{SUMCON}.

For calculating the feedback signal to check the actuator range, we have to design the open-loop transfer functions of the length control loops. \hyou{LengthUGF} summarizes the unity-gain frequencies of each loop. The unity-gain frequencies are set to 200~Hz for DARM and 50~Hz for all the other degrees of freedom. The cross-over frequency between the OP stage loop and the IM stage loop is set to 10~Hz, and that between the IM stage and the MN stage is set to 5~Hz. The IMCL control system comprises two loops. One is the MCe position control loop and the other is the frequency actuator loop used to stabilize the laser frequency. In this paper, we only consider the MCe position control loop for the sake of simplicity.

\begin{table}
  \begin{center}
    \caption{Mirrors used for the feedback, and unity-gain frequencies (UGFs) of the mirror loops, for the length control assumed in the modeling.} \label{LengthUGF}
    \begin{tabular}{lccccc}
      \br
     & Mirror(s) & UGF \\
\hline
DARM (differential arm length)        & ETMs & 200 Hz \\
MICH (Michelson differential length)  & BS   & 50 Hz \\
PRCL (power recycling cavity length) & PRM  & 50 Hz \\
SRCL (signal recycling cavity length)& SRM  & 50 Hz \\
IMCL (input mode cleaner length)     & MCe  & 50 Hz \\
      \br
    \end{tabular}
  \end{center}
\end{table}

The noises included in this model are described in the following subsections. The contribution from each noise should be below the displacement noise requirement for each mirror. Displacement noises of Type-A mirrors define the KAGRA sensitivity, and the requirement is set so that all the technical noises are below the thermal noise and the quantum noise limit. The Type-B and Type-Bp mirrors also have displacement noise requirements since there is a non-negligible coupling of auxiliary degrees of freedom to the DARM signal. IMC mirrors have a displacement noise requirement since the IMC is used for the laser frequency stabilization, as discussed in \Sec{Sec:IFO}. 

To derive the displacement noise requirements, we calculated the transfer functions from displacements of the mirrors and laser frequency fluctuations to the DARM signal using Optickle. The displacement noise requirements were set so that the effects of the mirror displacements on the DARM signal are smaller than the DARM noise above 10~Hz, which is the lower end of the observation band. Details of the requirements can be found in Refs.~\cite{AsoKAGRA,MIFDesign}. 

It should be noted that the cryocooler vibration coupled via heat links is not included in our model. The heat links are made of high-purity thin aluminum wires and they are soft enough that they do not significantly change the mechanical response of the suspension. The wiring of the heat links are designed so that the vibration from the cryocoolers does not affect the KAGRA sensitivity~\cite{DanVibration}. The effect from the heat links are not significant for the actuation design compared with the noises described below~\cite{SekiguchiPhD}.

\subsection{Seismic noise}

\begin{figure}
	\begin{center}
		\includegraphics[width=12cm]{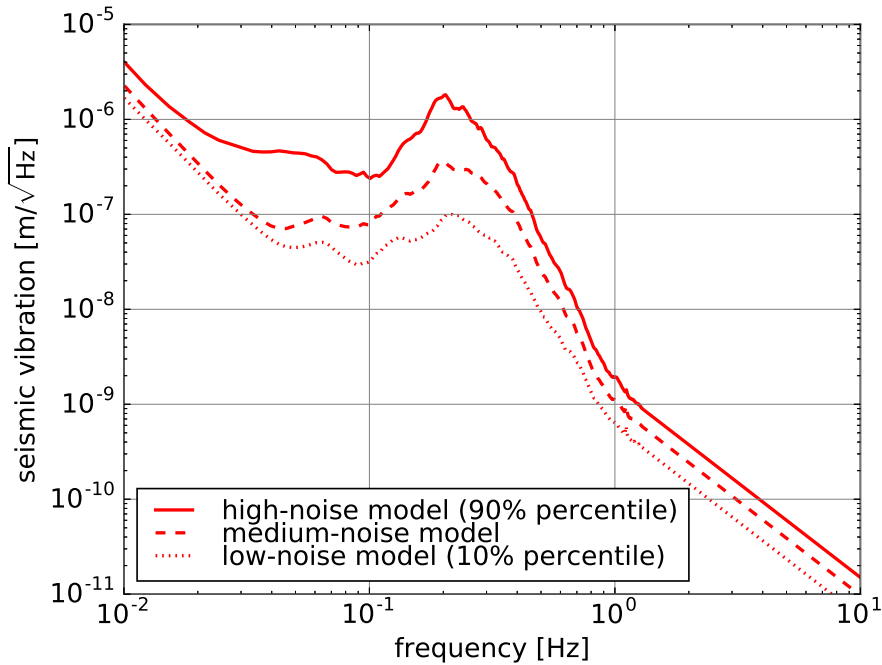}
	\caption{Ground seismic vibration models in the Kamioka mine based on a 1.5-year long measurement. The high-noise model spectrum was used in our model.}
	\label{KamiokaSeismic}
	\end{center}
\end{figure}

The motion of the mirror without any feedback is mostly determined by the seismic vibration of the ground. To estimate the seismic noise of the mirrors, we used the simulated vibration isolation ratio using SUMCON and the measured spectrum of the seismic vibration in the Kamioka mine.

The ground seismic vibration data used in our simulation was taken between September 2009 and February 2011 with a Guralp CMG-3T seismometer. Using this 1.5-year long data, we have created low-noise, medium-noise and high-noise models of the Kamioka mine from 10\% percentile, median and 90\% percentile spectra, respectively~\cite{SekiguchiSeismic}.
Figure~\ref{KamiokaSeismic} shows the models of the seismic vibration in the Kamioka mine. Since the measured data was limited by the sensor noise above 1.5~Hz, we assumed $1/f^2$ spectra above that frequency~\cite{Seismology}. The seismic vibration varies by roughly an order of magnitude depending on the weather, wind speed, human activities, etc. The peak at around 0.2~Hz comes from ocean waves, and is called the microseismic peak. To be conservative, we used the high-noise model spectrum for our simulation.

Our simulation focuses on the longitudinal degrees of freedom of mirrors. However, vertical vibration of a mirror also couples into the interferometer length signal due to unintentional asymmetry of the suspension system and slight tilt of the laser beam axis. The KAGRA tunnel and the laser beam axis are intentionally tilted by 1/300 to naturally drain water inside the mine~\cite{UchiyamaTunnel}. This vertical to longitudinal coupling is critical since the vibration isolation ratio for the vertical motion is lower. In our simulation, we assumed 1\% for this coupling factor.

\subsection{Magnetic noise}
The actuation efficiency of a mirror is determined by the gain and the output resistance of the coil driver, the number of coil turns, and the magnetic moment of the magnet. The size of the magnet is related to the magnetic noise coupling and thus it must be sufficiently small to minimize the coupling. As depicted in \zu{SuspensionSystems}, every KAGRA suspension, except for the BS, has four magnets attached to the back surface of the mirror. The BS has four magnets attached to the front surface instead to ensure that the magnets do not interfere with the beam. At BS, the beam is centered on the front surface but not on the back surface. The polarizations of four magnets are alternated to cancel the net magnetic moment. However, there is a residual magnetic moment since the magnetic moment of the magnets are not perfectly equal. The residual magnetic moment couples with environmental magnetic field gradient fluctuation and this creates magnetic force noise acting on the mirror. This force noise can be estimated with
\begin{equation}
 F_{1} = 2 \delta \mu_{\rm mag} \partial_{\rm l}B ,
\end{equation}
where $\delta \mu_{\rm mag}$ is the variation of the magnetic moment and $\partial_{\rm l}B$ is the environmental magnetic field gradient in the longitudinal direction.

From measurements, $\delta \mu_{\rm mag}$ is estimated to be smaller than 20\% of the nominal magnetic moment.
For $\partial_{\rm l}B$, we used the estimated gradient from the typical magnetic field fluctuation measured inside the Kamioka mine with a magnetometer Phoenix Geophysics AMTC-30~\cite{AtsutaMagnetic}. Assuming the typical length scale for the magnetic field fluctuation to be 1~m as an pessimistic case~\cite{ConteMagnet}, $\partial_{\rm l}B$ is estimated to be $4 \times 10^{-13} \unit{T/m/\rtHz}$ at 10~Hz.

The environmental magnetic field gradient difference in the positions of four magnets also generates magnetic force noise. This can be estimated with
\begin{equation}
 F_{2} = 2 \mu_{\rm mag} \delta (\partial_{\rm l}B) ,
\end{equation}
where $\mu_{\rm mag}$ is the magnetic moment of a magnet and $\delta (\partial_{\rm l}B)$ is the magnetic field gradient difference. Measurements show that $\delta (\partial_{\rm l}B)$ is roughly 10\% of the nominal magnetic field gradient~\cite{ConteMagnet}.

There are other coupling mechanisms for the magnetic noise, but they are found to be negligible. The environmental magnetic field induces torque on the magnets, which leads to rotational motion of the mirror if the torques acting on four magnets do not perfectly cancel each other. This noise couples into the interferometer length signal when the beam spot is not perfectly centered on the mirror~\cite{Kawamura1994}. Assuming a worst case in which the beam mis-centering is 1~mm~\cite{AsoKAGRA} and magnetic field difference between four magnets to be 10\%, this effect is estimated to be roughly an order of magnitude smaller than the effect from $F_{1}$ and $F_{2}$.
The environmental magnetic field gradient also creates a lateral force acting on the magnets, which creates a torque on the mirror if forces acting on four magnets do not perfectly cancel each other. This effect is estimated to be two orders of magnitude smaller. Magnetization of the mirror itself also couples with the environmental magnetic field, but this force noise is estimated to be more than three orders of magnitude smaller. Therefore, we only included the force noise from $F_{1}$ and $F_{2}$ in our model.

Similar calculations were also done for the IM and MN stages. Since the force noises from the upper stages are attenuated by the suspension, IM and MN stages can have larger magnets than the OP stage, which gives larger actuation range at low frequencies below $\sim 10\unit{Hz}$. For Type-B and Type-Bp suspensions, of the six actuation magnets on the IM, only one is for longitudinal actuation. An extra magnet with opposite polarization is inserted inside a magnet holder, called a flag, to cancel the magnetic moment~\cite{MarkOSEM}.

\subsection{Electronics noise}
We can split the noises from various electronics into two categories, sensor electronics noises and actuator electronics noises. Sensor electronics noises are noises which come from the displacement sensing of the mirror, such as ADC noise, whitening filter noise, and dark current of the photo-diodes, and are not included in our model. Electronics for the sensing are designed so that their noises are smaller than the quantum noise described in the next subsection, and therefore, can be neglected from the calculation. The effect of the photo-diode noise is discussed in \Ref{AsoLSC}.

Here, we focus on the actuator electronics noises which couple into the loop after the servo filter. Noises of the actuation system creates force noise by fluctuation of the current applied to the coils. Contribution from these noises to the mirror displacement noise depends solely on the actuation design, and is independent of the servo filter design.

We included the calculated coil driver noise and the measured DAC noise in our simulation. The coil driver noise is calculated from operational amplifier noises and Johnson noise of resisters. The input-equivalent noises are $8 \times 10^{-9} \unit{V/\rtHz}$ and $2 \times 10^{-8} \unit{V/\rtHz}$ at 10~Hz for high power and low power coil driver, respectively.

The DAC noise is measured to be $2 \times 10^{-6} \unit{V/\rtHz}$ at 10~Hz and is the largest actuator electronics noise. To effectively reduce the DAC noise, we use switchable three-stage whitening filters and dewhitening filters, except for in the IMC suspensions. Each stage of the whitening filter has a zero at 1~Hz and a pole at 10~Hz, and vice versa for the dewhitening filters. In our simulation, all three stages are turned on, which means that the DAC noise is effectively reduced by three orders of magnitude above 10~Hz. Noise from the anti-aliasing and anti-imaging filters are measured to be $4 \times 10^{-8} \unit{V/\rtHz}$ at 10~Hz, and are negligible compared with the DAC noise.

\subsection{Quantum noise} \label{Sec:Quantum}
The displacement sensing of a mirror with an interferometer is ultimately limited by quantum noise. The quantum fluctuation of the laser power creates force noise on the mirror and this results in quantum radiation pressure noise. The quantum fluctuation of the laser power on the photo-diode creates shot noise. The shot noise dominates the sensing noise at high frequency, and this creates excess force noise via the control loop. Since the actuation efficiency is limited at high frequency, the shot noise contribution is an important factor for the actuation range. In our simulation, we again used Optickle to calculate the quantum noise for each length degree of freedom.

As for the DARM loop, we ran the simulation with the quantum noise from the main 1064~nm laser and also with the quantum noise from the arm-length stabilization system to check for the range of the actuator. The shot noise of the length sensing of the arm cavity with auxiliary 532~nm laser is worse than that for the main laser because the input power and the finesse are less. The input laser power at 532~nm to the arm cavities is assumed to be 100~mW in our model.

\section{Results}

\def\miniwid{6.5cm}
\begin{figure}
	\begin{center}
\begin{minipage}[b]{0.49\textwidth}
   \begin{center}
   \includegraphics[width=\miniwid]{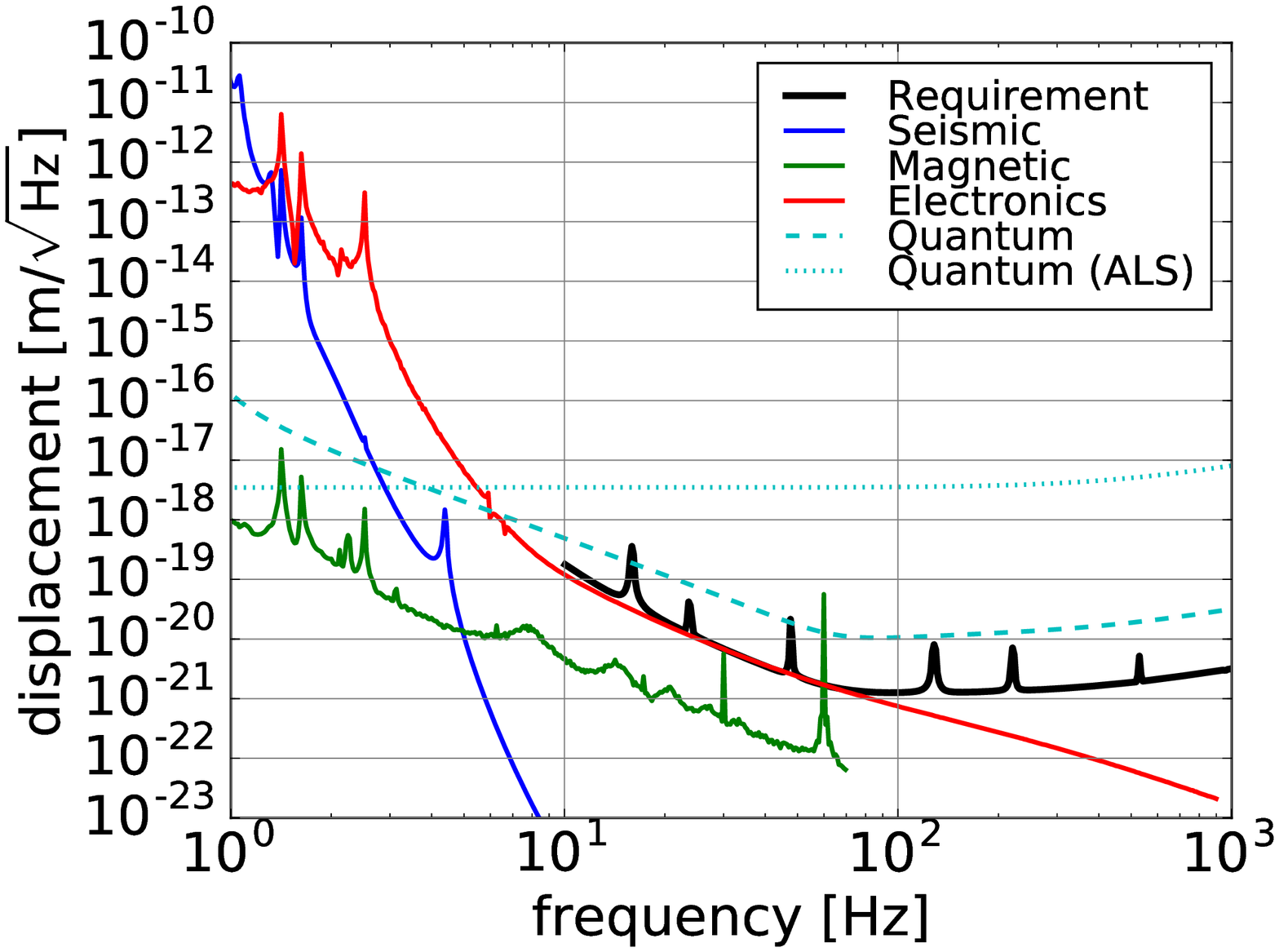} \\
   (a) Type-A (ETM) \\
   \includegraphics[width=\miniwid]{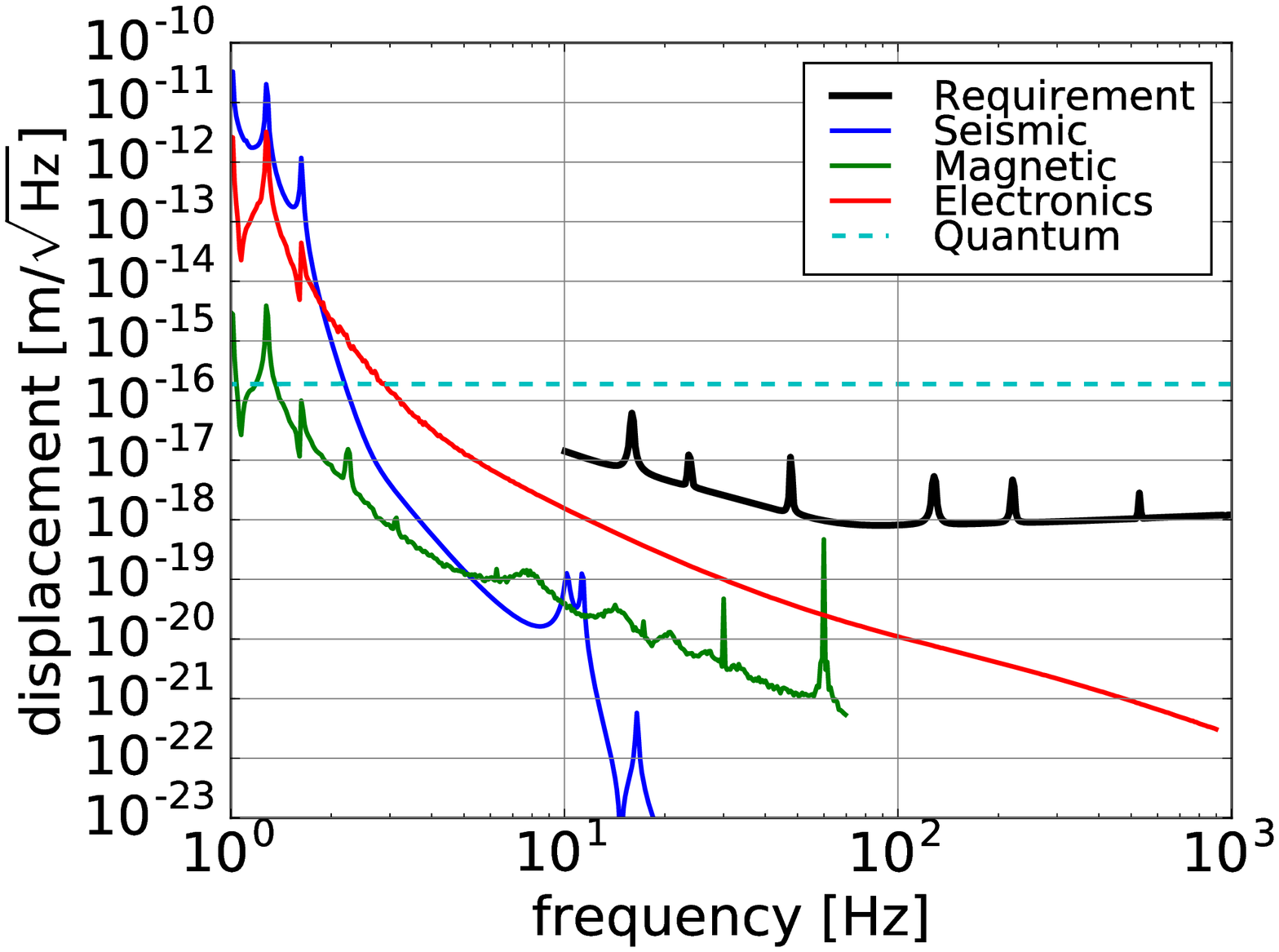} \\
   (b) Type-B (BS) \\
   \includegraphics[width=\miniwid]{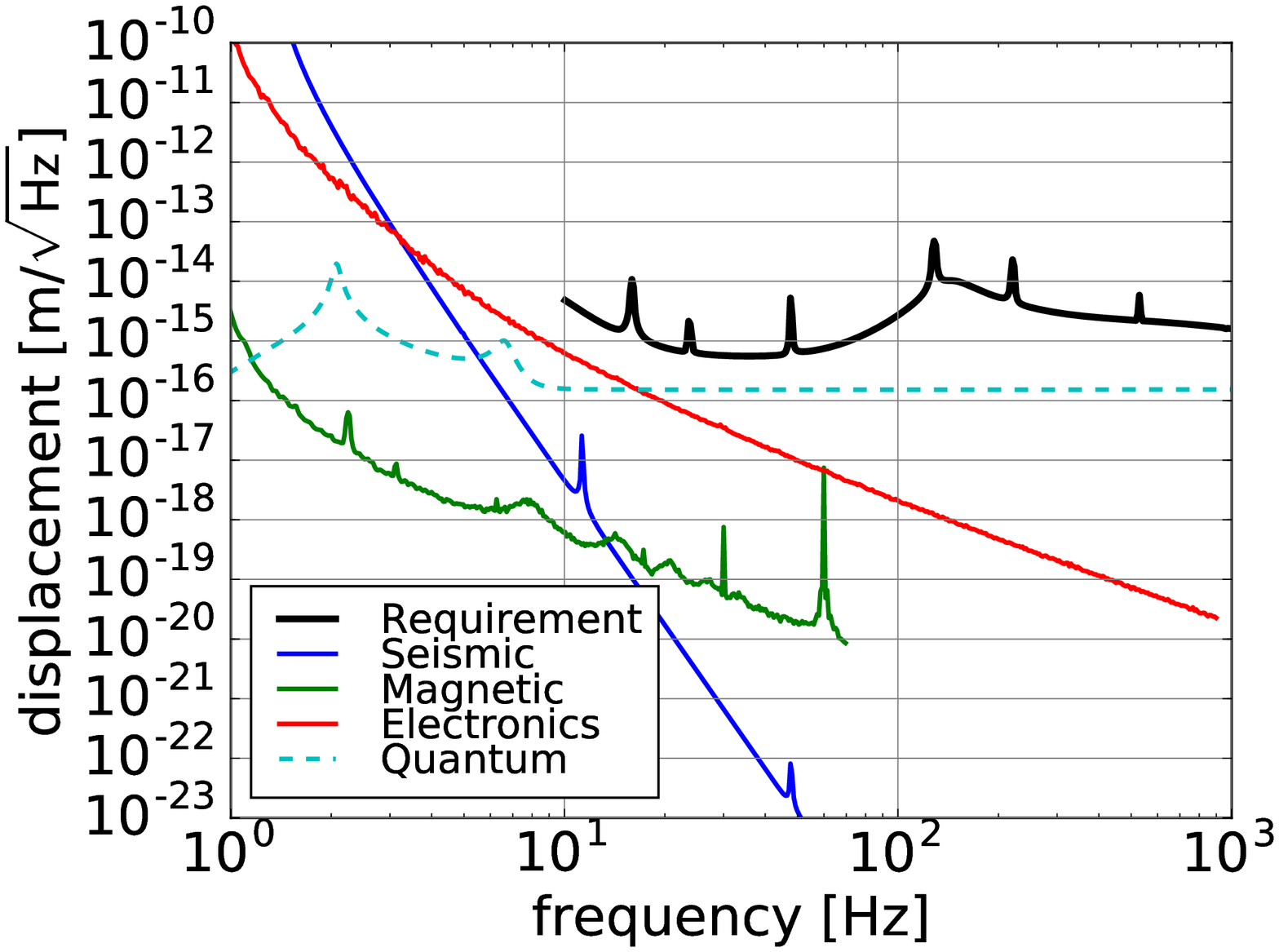} \\
   (d) Type-Bp (PRM)
   \end{center}
\end{minipage}   
\begin{minipage}[b]{0.49\textwidth}
   \begin{center}
   \includegraphics[width=\miniwid]{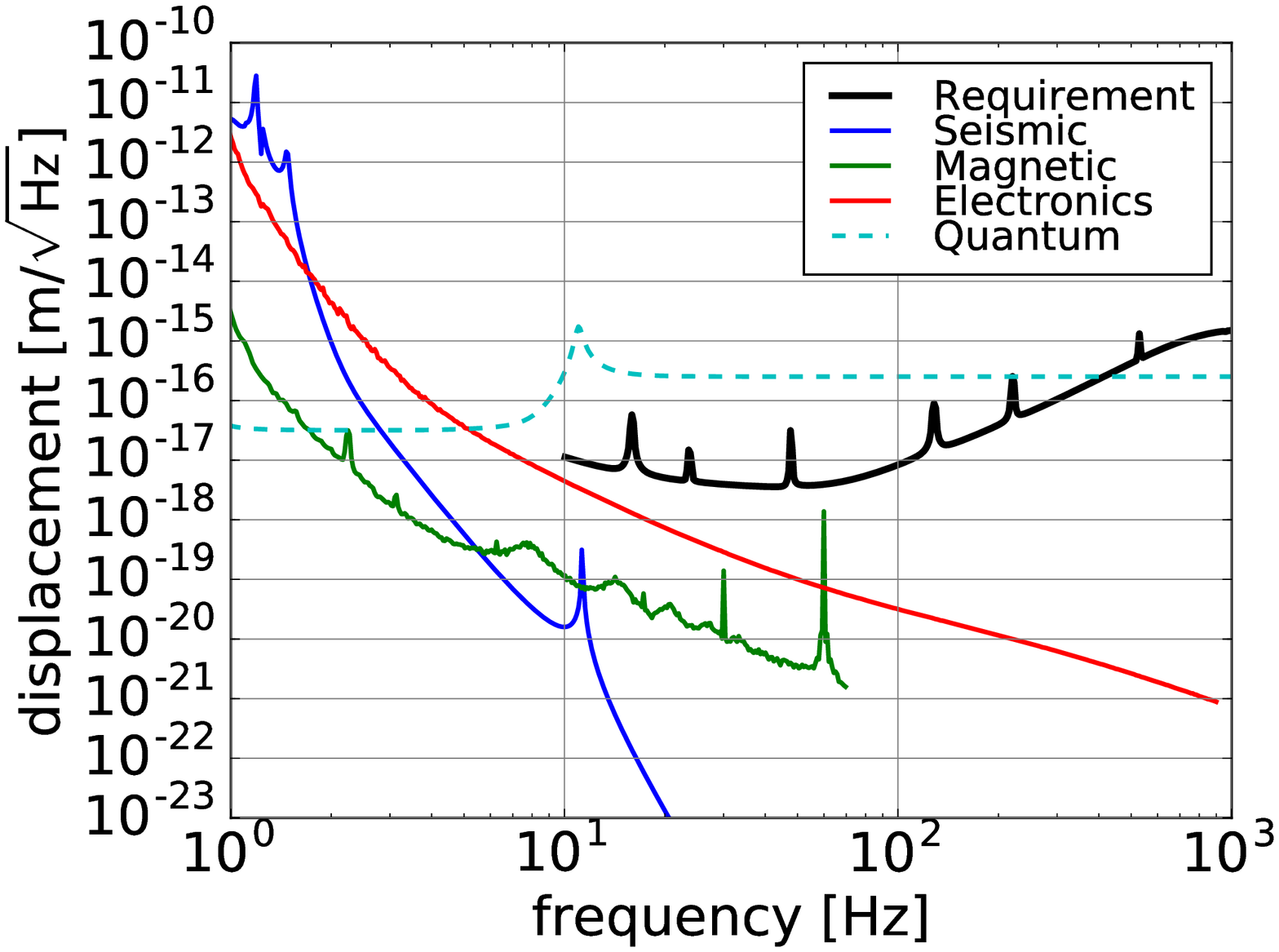} \\
   (c) Type-B (SRM) \\
   \includegraphics[width=\miniwid]{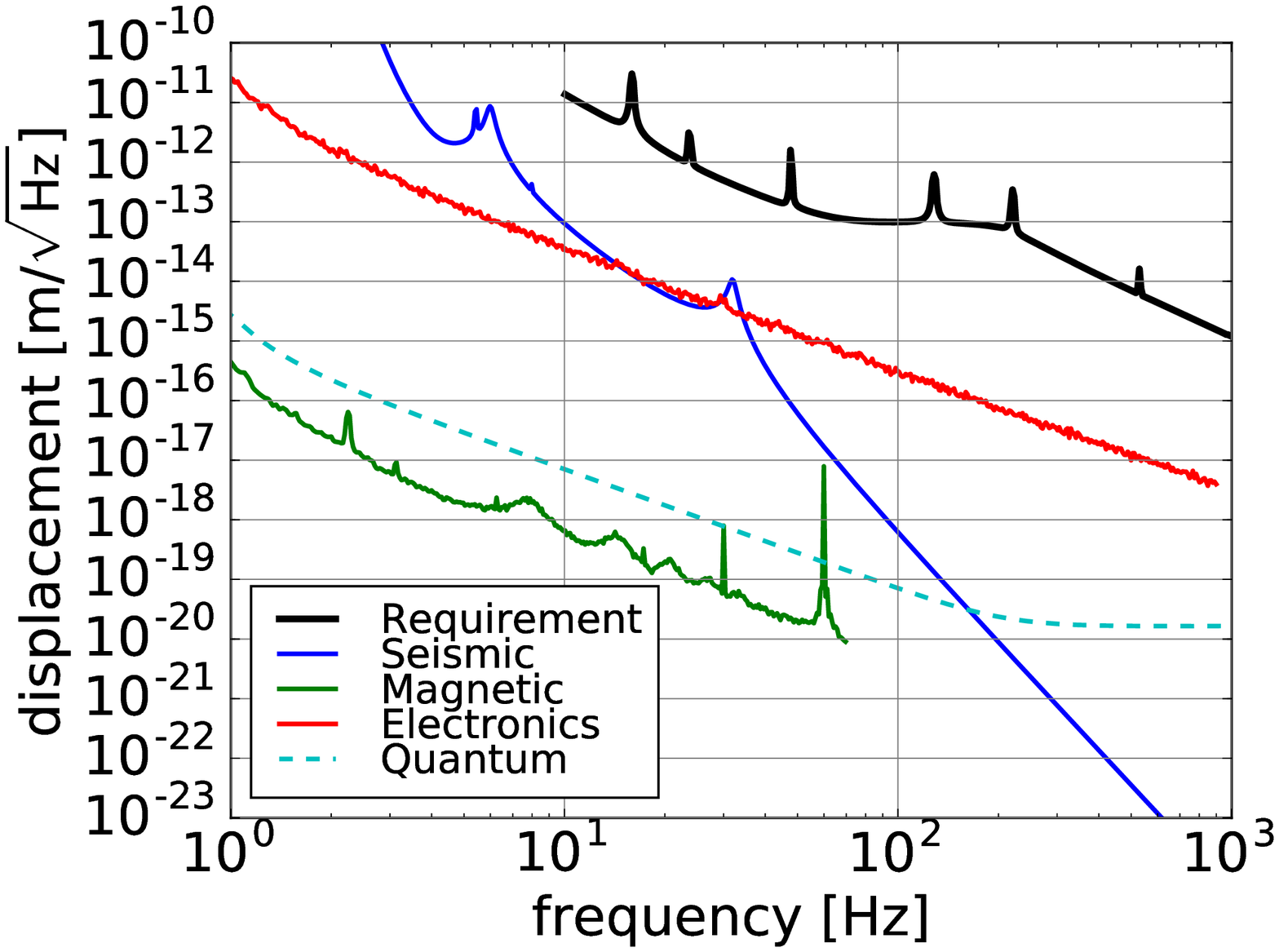} \\
   (e) Type-C (MCe)
   \end{center}
\end{minipage}
	\caption{Simulated displacement noise spectra of the mirrors. The thick black lines show the displacement noise requirement with a safety factor of 10. For Type-A, quantum noise in the arm-length stabilization (ALS) system is also plotted.} \label{DisplacementNoise}
	\end{center}
\end{figure}

The simulated displacement noise spectra of the mirrors using our model are shown in \zu{DisplacementNoise}. The displacement noise requirement plotted in the figure has a safety factor of 10 to ensure that technical noises other than the quantum noise are sufficiently small. The seismic noise is smaller than the requirement, confirming that the KAGRA vibration isolation systems sufficiently suppress the ground vibration. The magnetic noise and the electronics noise are also sufficiently small. The peaks at 30~Hz and 60~Hz in the magnetic noise spectra are from AC power lines. For every suspension except for IMC, the electronics noise above $\sim 5\unit{Hz}$ is dominated by the coil driver noise from the OP stage. For IMC, electronics noise is dominated by DAC noise, since no whitening and dewhitening filters are applied.

The quantum noise for the Type-A suspension is 10 times larger than the requirement with a safety factor of 10. This is because the requirement is set from the DARM displacement sensitivity, which defines the KAGRA sensitivity to gravitational waves, and the quantum noise limits the sensitivity above $\sim 40 \unit{Hz}$. The DARM quantum noise in the arm-length stabilization system is also plotted to show that the shot noise is more than two orders of magnitude larger than that for the main laser.

\begin{figure}
	\begin{center}
\begin{minipage}[b]{0.49\textwidth}
   \begin{center}
   \includegraphics[width=\miniwid]{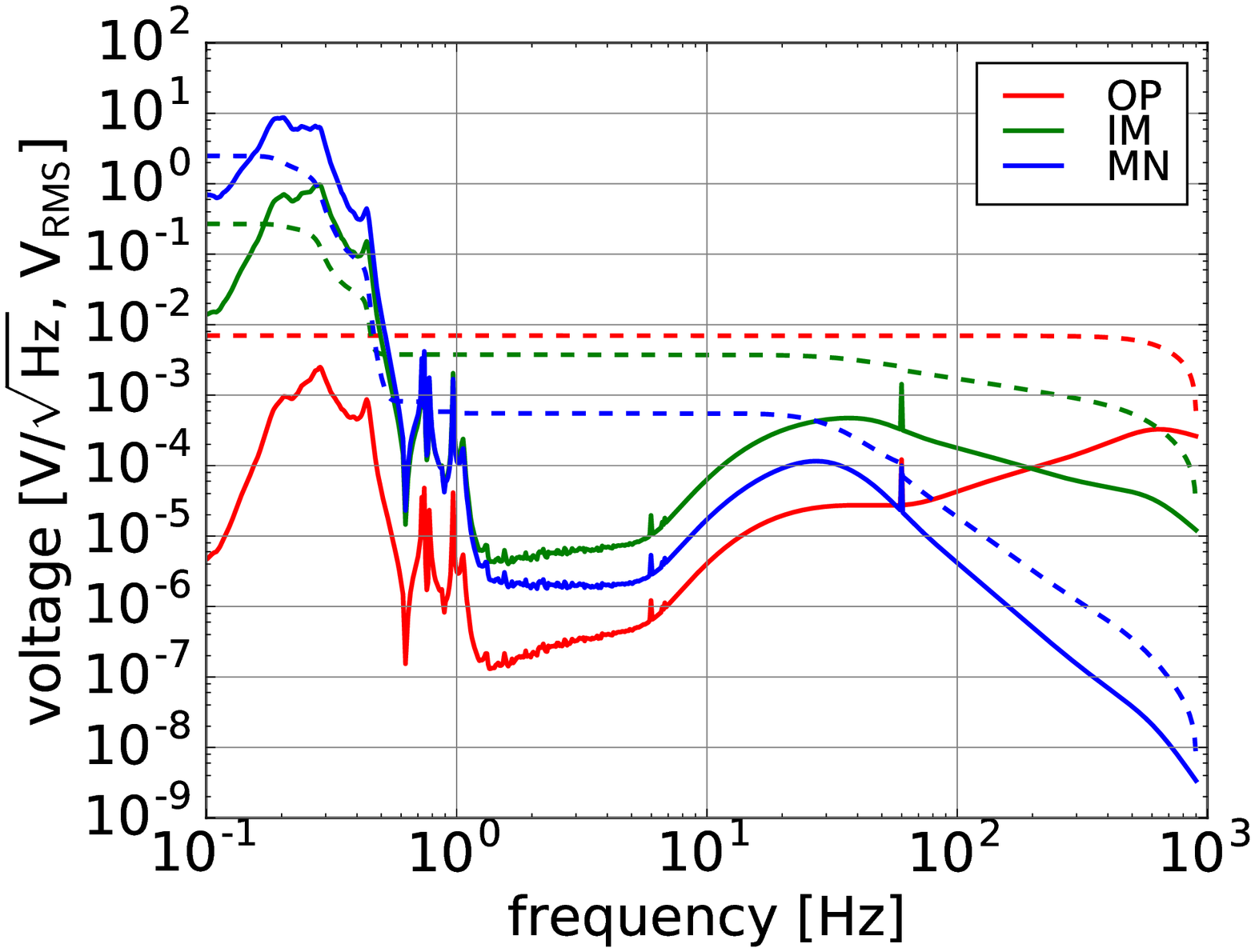} \\
   (a) Type-A (ETM) \\
   \includegraphics[width=\miniwid]{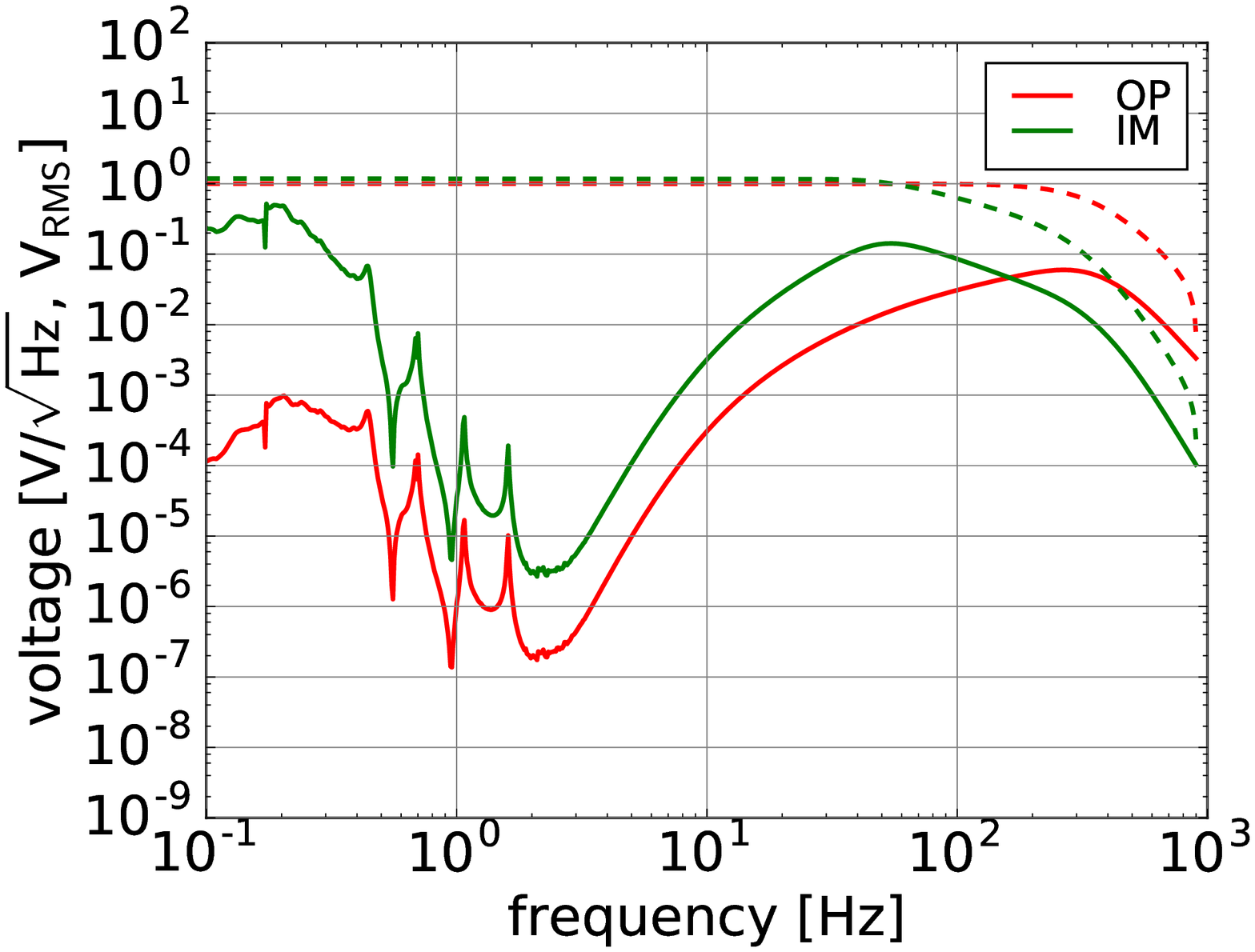} \\
   (b) Type-B (BS) \\
   \includegraphics[width=\miniwid]{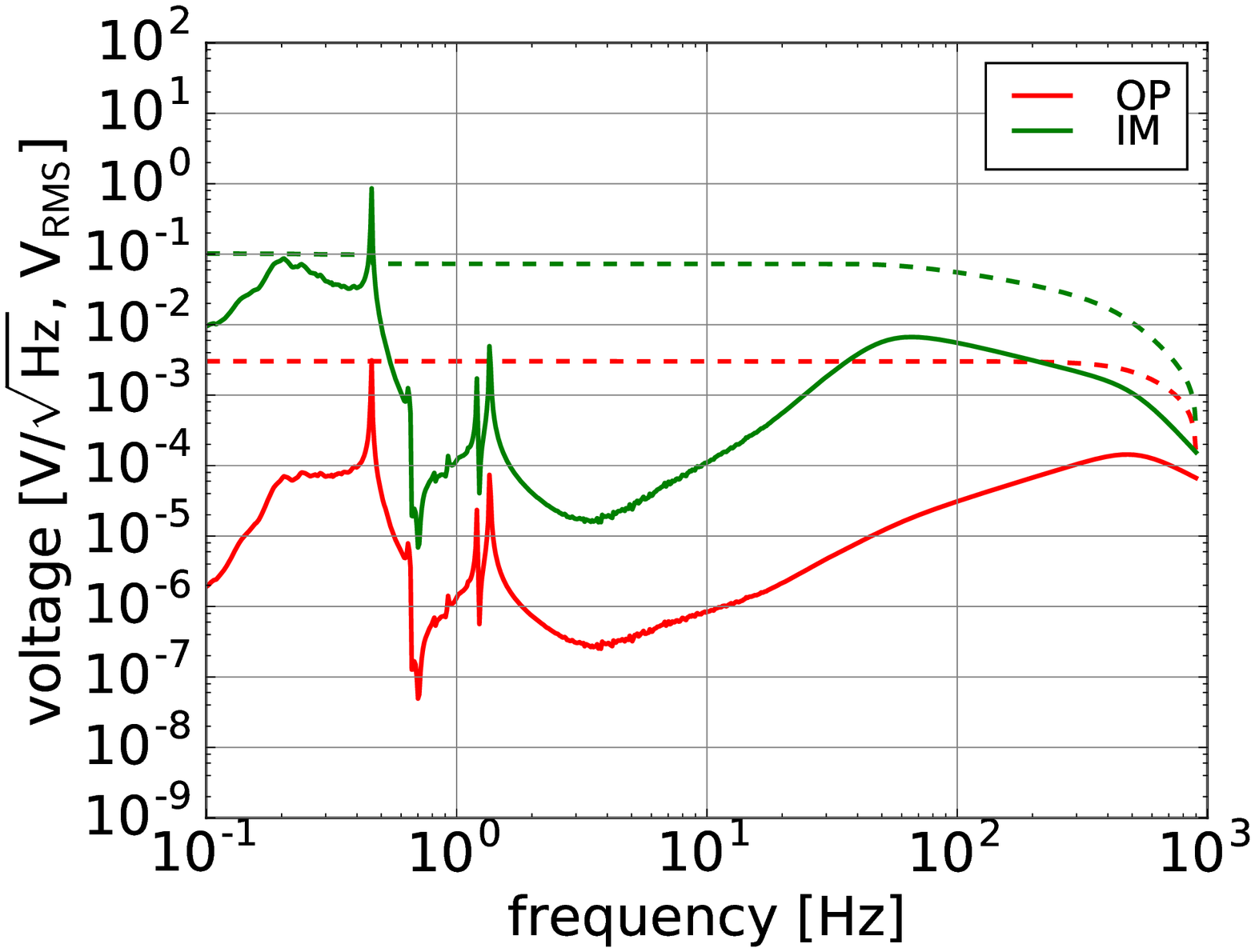} \\
   (d) Type-Bp (PRM)
   \end{center}
\end{minipage}   
\begin{minipage}[b]{0.49\textwidth}
   \begin{center}
   \includegraphics[width=\miniwid]{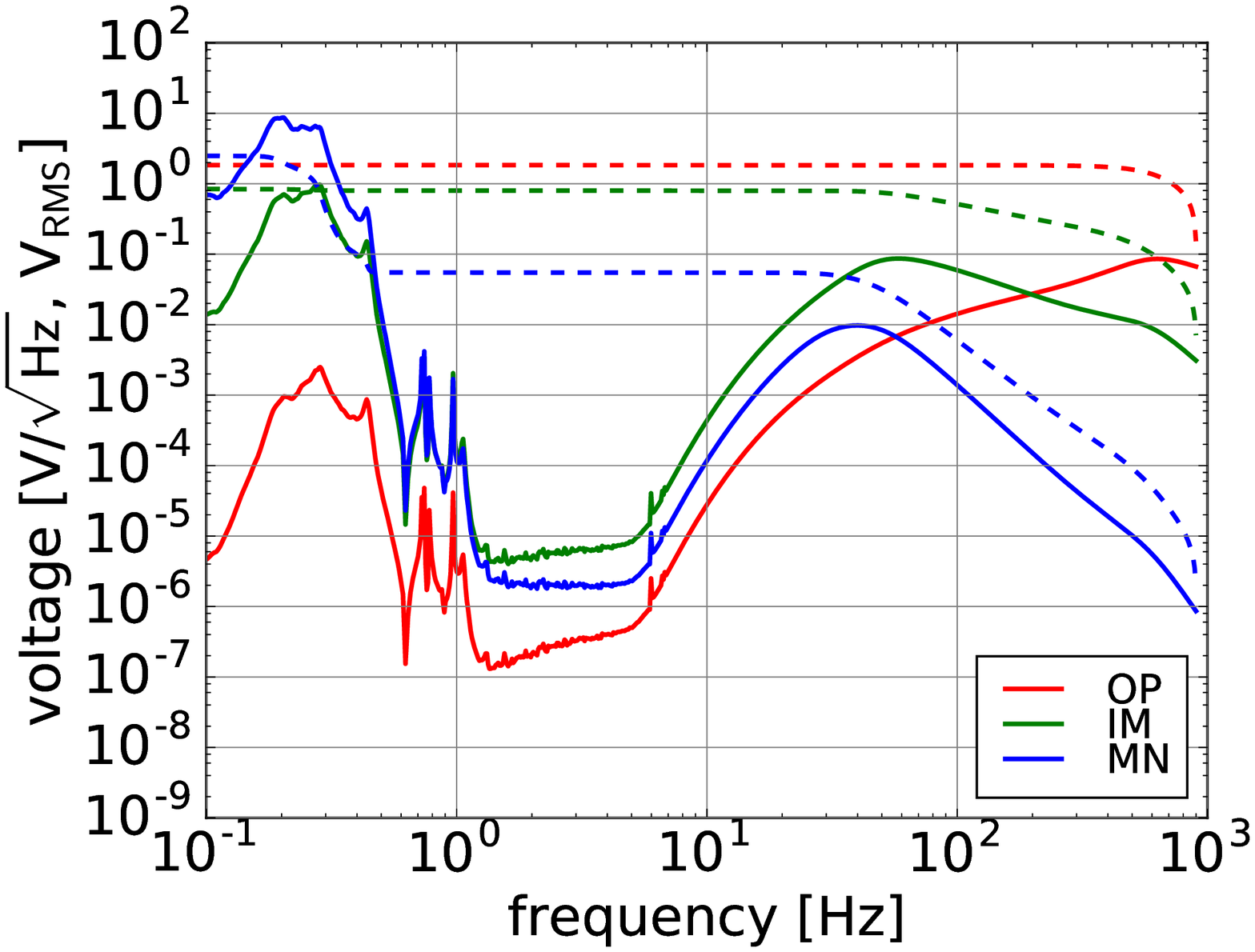} \\
   (a') Type-A (ETM) with ALS \\
   \includegraphics[width=\miniwid]{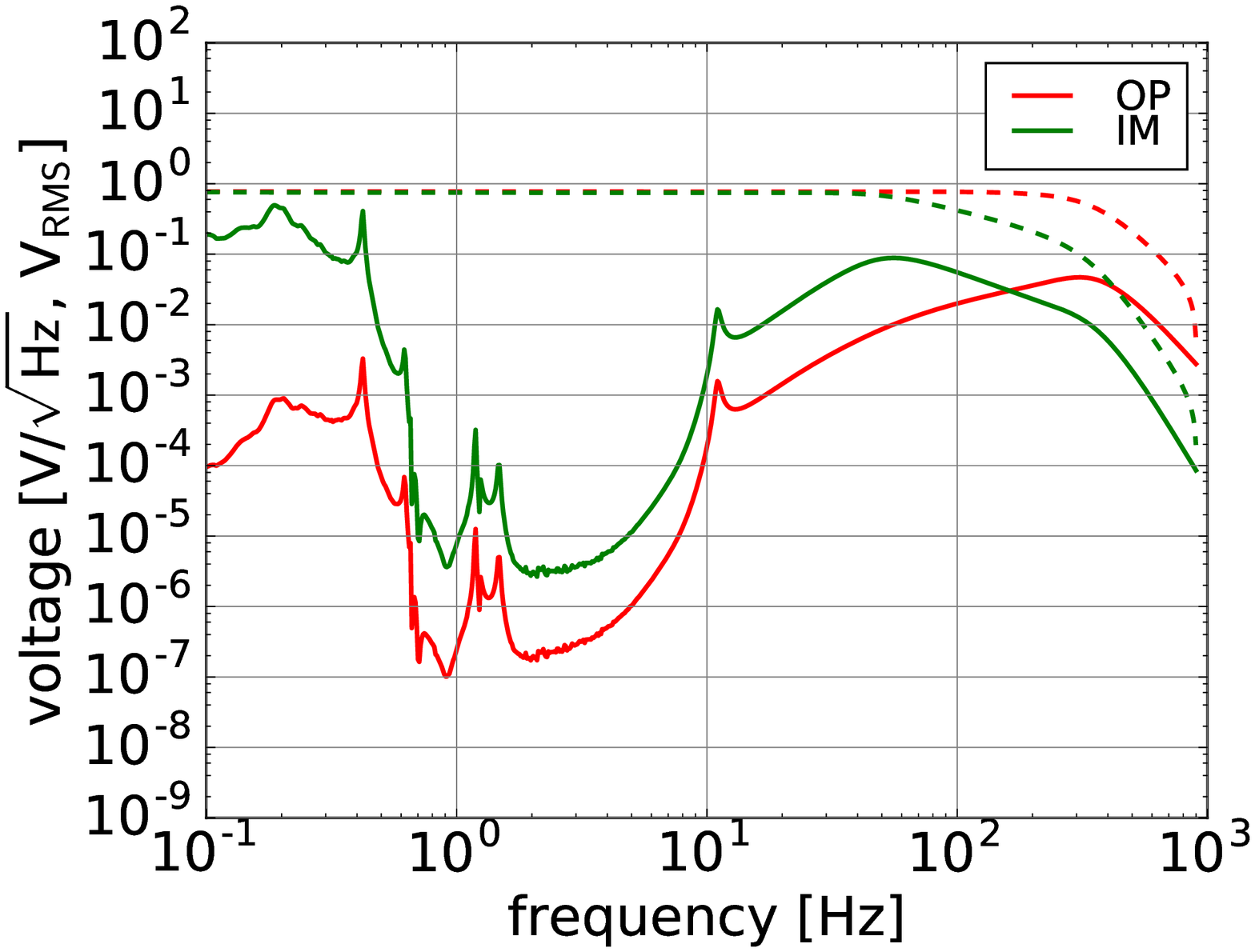} \\
   (c) Type-B (SRM) \\
   \includegraphics[width=\miniwid]{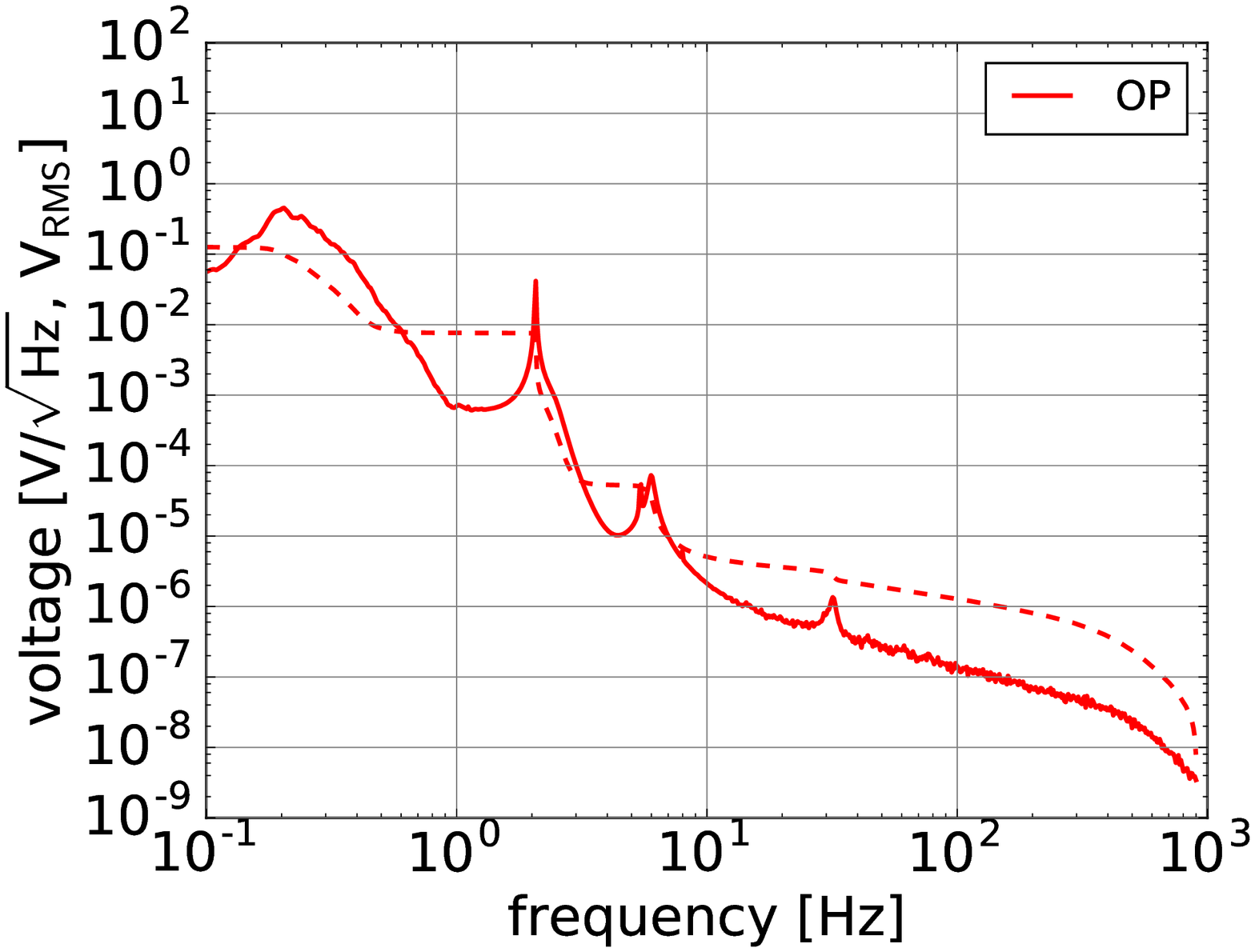} \\
   (e) Type-C (MCe)
   \end{center}
\end{minipage}
	\caption{Spectra of simulated feedback signals for suspension stages. Cumulative RMS is plotted with dotted lines in units of volts. It is required that the RMS does not exceed the DAC range of 10 V. For Type-A, spectra when the arm-length stabilization system is used are also plotted in (a').} \label{Feedback}
	\end{center}
\end{figure}

The quantum noise for Type-B suspensions, however, has to be mitigated. We will cancel the effect of this on the DARM signal with a feed-forward technique~\cite{AsoKAGRA,AsoLSC}. Using transfer functions from the actuation of the auxiliary degrees of freedom to the DARM signal, we can cancel the effect from the excess mirror displacement caused by the auxiliary length control loops. Our calculation show that a feed-forward gain of 100 is sufficient to mitigate this quantum noise coupling. The feed-forward gain of 100 corresponds to a transfer function estimation accuracy of 1\%, and this is technically feasible.

The spectra of simulated feedback signals for each suspension stage are shown in \zu{Feedback}. The root-mean-squares (RMSs) of the feedback signals are 2~V at maximum and do not exceed the DAC range of 10~V, which means that the actuator strengths are sufficiently large. For every suspension except for the IMC, the lower frequency part of the feedback signal is dominated by seismic noise, and the higher frequency part is dominated by quantum noise. For IMC, the higher frequency part is dominated by electronics noise.

For the Type-A suspension, the spectrum when the arm-length stabilization system is used is also plotted in \zu{Feedback} (a'). The quantum noise at higher frequency is large and uses up most of the actuation range for the OP and IM stages, but nonetheless, the RMS does not exceed the DAC range.

The calculated actuator force range and actuator noise for each suspension at each stage are summarized in \hyou{DesignSummary}.

\begin{table}
  \begin{center}
    \caption{Summary of KAGRA mirror actuation design. Maximum force actuator can produce with DAC output of 10~V and actuation efficiency at DC, and sum of actuator electronics and magnetic noises at 10~Hz for each suspension at each stage are shown.} \label{DesignSummary}
    \begin{tabular}{ccccc}
      \br
 & & Max force [N] & Efficiency [m/V] & Noise [$\rm{m/\rtHz}$] \\
\hline
Type-A      & OP & $7.7 \times 10^{-6}$ & $1.8 \times 10^{-9}$ & $1.0 \times 10^{-19}$ \\
            & IM & $1.5 \times 10^{-4}$ & $1.7 \times 10^{-8}$ & $4.4 \times 10^{-20}$ \\
            & MN & $8.2 \times 10^{-3}$ & $3.9 \times 10^{-7}$ & $3.6 \times 10^{-20}$ \\
Type-B (BS) & OP & $9.4 \times 10^{-5}$ & $2.6 \times 10^{-8}$ & $1.5 \times 10^{-18}$ \\
            & IM & $2.1 \times 10^{-3}$ & $1.6 \times 10^{-7}$ & $1.7 \times 10^{-19}$ \\
Type-B (SR) & OP & $1.5 \times 10^{-4}$ & $7.3 \times 10^{-8}$ & $4.4 \times 10^{-18}$ \\
            & IM & $2.1 \times 10^{-3}$ & $4.2 \times 10^{-7}$ & $4.2 \times 10^{-19}$ \\
Type-Bp     & OP & $5.7 \times 10^{-2}$ & $2.7 \times 10^{-5}$ & $6.2 \times 10^{-16}$ \\
            & IM & $1.4 \times 10^{-1}$ & $2.6 \times 10^{-5}$ & $1.0 \times 10^{-17}$ \\
Type-C      & OP & $7.1 \times 10^{-4}$ & $4.3 \times 10^{-6}$ & $3.6 \times 10^{-14}$ \\
      \br
    \end{tabular}
  \end{center}
\end{table}

\section{Actuation range for lock acquisition}
Our calculation is done in the frequency-domain, and our model is based on the static response of the interferometer. Feedback signals when acquiring the lock of the interferometer should also be calculated to check the force range of actuators.

The force we need to stop the mirror can be roughly estimated from the relationship between the mirror momentum and the impulse,
\begin{equation}
  F = \frac{m v}{\Delta t},
\end{equation}
where $m$ and $v$ are the mass and the velocity of the mirror, respectively. $\Delta t$ is the time it takes to pass the linewidth $d$ of the cavity. Since $\Delta t = d / v$, the velocity requirement for the mirror can be calculated as
\begin{equation}
  v_{\rm req} = \sqrt{\frac{F_{\rm max} d}{m}}.
\end{equation}
For a Michelson interferometer, $d$ can be estimated with half of the laser wavelength.

Assuming all the feedback is done at the OP stage during lock acquisition, we can set $F_{\rm max}$ to be the maximum force the OP actuator can produce (see \hyou{DesignSummary}). Calculated velocity requirements for BS, SRM, PRM, and MCe are $1.6\unit{\mu m/sec}$, $0.44\unit{\mu m/sec}$, $7.3\unit{\mu m/sec}$, and $1.2\unit{\mu m/sec}$, respectively. Suspension modeling using SUMCON shows that these requirements can be fulfilled at the medium-noise seismic level (see \zu{KamiokaSeismic}), with sufficient damping servo~\cite{SekiguchiPhD}.

For the Type-A suspension, lock acquisition is done with the laser frequency actuator of the arm-length stabilization system, with the feedback point then gradually changed from the frequency actuator to the mirror actuator. Therefore, the discussion above cannot be applied. As for the frequency actuator, we use an acousto-optic modulator which have been confirmed to have enough range. Details of the arm-length stabilization system and the laser frequency stabilization system are beyond the scope of this paper and will be discussed elsewhere.

\section{Conclusions}
We have developed a frequency-domain model, which incorporates various noises and realistic suspension responses, to simulate the length control of the KAGRA interferometer using mirror actuators. With our model, we have determined the mirror actuation parameters which give sufficient range to control all the interferometer length degrees of freedom, while keeping the displacement noise from the actuation sufficiently small. We have also checked that the actuators have sufficient range for acquiring the lock of the interferometer.

Minimizing the effect of controls noise is important for realizing the target sensitivity at the lower end of the detector bandwidth. Our model can also be used to design the mirror actuators of future interferometric gravitational wave telescopes.

\ack
We would like to thank Ettore Majorana, Jeffrey Kissel, Christopher Wipf and Stefan Ballmer for fruitful discussions. We appreciate the technical support from the Advanced Technology Center (ATC) of NAOJ.
T. S., Y. E., K. K. and H. T. acknowledges financial support received from the Advanced Leading Graduate Course for Photon Science (ALPS) program at the University of Tokyo.
C. P. O. acknowledges financial support received from the Global Science Graduate Course (GSGC) at the University of Tokyo.

The KAGRA project is supported by MEXT, JSPS Leading-edge Research Infrastructure Program, JSPS Grant-in-Aid for Specially Promoted Research 26000005, MEXT Grant-in-Aid for Scientific Research on Innovative Areas 24103005, JSPS Core-to-Core Program, A. Advanced Research Networks, the joint research program of the Institute for Cosmic Ray Research, University of Tokyo, National Research Foundation (NRF) and Computing Infrastructure Project of KISTI-GSDC in Korea, the LIGO project, and the Virgo project.

\section*{References}
\bibliographystyle{iopart-num}
\providecommand{\newblock}{}

\end{document}